\documentstyle[12pt,aaspp4]{article}

\def\hour{\hbox{$^{\rm h}$}}
\def\icm{${\rm cm}^{-1}$}
\def\he3{$^3{\rm He}$}
\def\COBE{{\sl COBE}}
\def\IRAS{{\sl IRAS}}
\def\bx{\mbox{\boldmath $x$}}
\def\rawi{\hbox{${t_i}$}}
\def\rawj{\hbox{${t_j}$}}
\def\beami{\hbox{${B_i(\alpha)}$}}
\def\beamj{\hbox{${B_j(\alpha)}$}}
\def\covar{\hbox{${V_{ij}}$}}
\def\wgt{\hbox{${V^{-1}_{ij}}$}}
\def\fwhm{\hbox{$\theta_{\rm FWHM}$}}

\def\beq{\begin{equation}}
\def\eeq{\end{equation}}

\begin{document}

\slugcomment{Submitted to {\em Ap. J. Letters}; \quad {\tt astro-ph/9604155} }

\title{A Detection of Bright Features in the Microwave Background}

\author{
M.~S.~Kowitt\altaffilmark{1,2},
E.~S.~Cheng\altaffilmark{1},
D.~A.~Cottingham\altaffilmark{3},
D.~J.~Fixsen\altaffilmark{4},
C.~A.~Inman\altaffilmark{2,1},
S.~S.~Meyer\altaffilmark{2},
L.~A.~Page\altaffilmark{5},
J.~L.~Puchalla\altaffilmark{2,1},
J.~E.~Ruhl\altaffilmark{2,6},
and~R.~F.~Silverberg\altaffilmark{1}}
\altaffiltext{1}{Laboratory for Astronomy and Solar Physics,
     NASA/Goddard Space Flight Center, Code 685.0, Greenbelt, MD 20771}
\altaffiltext{2}{Enrico Fermi Institute, University of Chicago, 
     5640 South Ellis Avenue, Chicago, IL 60637}
\altaffiltext{3}{Global Science and Technology, Inc., 
     Laboratory for Astronomy and Solar Physics,
     NASA/GSFC Code 685.0, Greenbelt, MD 20771}
\altaffiltext{4}{Applied Research Corporation, 
     Laboratory for Astronomy and Solar Physics,
     NASA/GSFC Code 685.3, Greenbelt, MD 20771}
\altaffiltext{5}{Department of Physics, Princeton University, 
     Princeton, NJ 08544}
\altaffiltext{6}{Department of Physics, University of California, 
     Santa Barbara, CA  93106}

\begin{abstract} 

We report the characterization of bright, compact features in the 
cosmic microwave background radiation (CMBR) detected during the 
June 1992 and June 1994 balloon flights of 
the Medium Scale Anisotropy Measurement (MSAM1-92 and MSAM1-94, respectively). 
Spectral flux densities are determined for each feature at 
5.7, 9.3, and 16.5~\icm.
No viable counterparts for these features were found in source
catalogs at 5~GHz or at 100 \micron.
The measured spectrum of each feature is consistent with a temperature 
fluctuation in the CMBR. 
The existence of these features is consistent with
adiabatic fluctuation models of anisotropy in the CMBR.

\end{abstract} 

\setcounter{footnote}{0}

\section{Introduction}

Recent observations of the cosmic microwave background radiation (CMBR) have 
found statistically significant anisotropy at 0\fdg5 angular scales 
(see \cite{white94}, and references therein).
In earlier Letters (\cite{cheng94}, hereafter Paper~I, and \cite{cheng95},
hereafter Paper~II), 
we reported on observations of CMBR anisotropy that included two 
particularly bright features.
The presence of bright, unresolved sources suggested the possibility of
foreground point-source contamination, or of non-Gaussian fluctuations
in the CMBR.
In this Letter, we present our analysis of these features and 
a discussion of other work studying their properties.

\section{Instrument and Observations}

The MSAM1 instrument has been described in detail elsewhere (\cite{fixsen96a});
only the most relevant features are given here.
MSAM1 is a balloon-borne off-axis Cassegrain telescope with a 
nutating secondary mirror and a four-band bolometric radiometer. The 
detectors have 
center frequencies of 5.7, 9.3, 16.5, 
and 22.6 \icm\ each with $\sim 1.5$ \icm\ bandwidth. The beam size of the 
telescope is 28\arcmin\ FWHM, roughly independent of frequency. 
The secondary executes a horizontal 
four-position square-wave chop (i.e., center, left, center, right) with an 
amplitude of $\pm40\arcmin$ on the sky. The telescope is pointed 
using a combination of torque motors and reaction wheels, referenced to a 
gyroscope. The absolute sky location is determined by star camera 
observation. 

The data discussed here were collected during two flights of the MSAM
instrument from Palestine, Texas; the first on 1992~June~5 (MSAM1-92), 
the second on 1994~June~2 (MSAM1-94).  These observations have been described 
in Papers I and II, and are only summarized here.
In each flight, the antenna patterns for all four bands were measured 
by rastering over Jupiter.  We determine the location of the 
antenna pattern within the star camera's field-of-view to within 2\arcmin\ 
(post-flight) by simultaneously observing Jupiter in the star camera and 
the main telescope.  Gyroscope drift increases our absolute pointing 
uncertainty to $\pm 2\farcm5$.  
We calibrate our spectral flux density measurements  
using the measured surface temperatures (\cite{griffin86})
together with Jupiter's apparent diameter at the epoch of flight
(\cite{almanac92}, \cite{almanac94}).
The calibration has a 10\% overall uncertainty, with relative errors of
1--4\% between bands.

The CMBR observations were made at roughly constant declination 
$\delta \approx 82\arcdeg$, with the central beam covering right ascensions
14\fh4--16\fh9 and 17\fh2--20\fh3 during the MSAM1-92 flight.
This corresponds to $\sim 6.6$ square degrees of sky coverage.
The goal of the MSAM1-94 flight was to observe the identical field as MSAM1-92;
this was nearly achieved, with an average declination difference of 
$\sim 10\arcmin$.  The coverage in right ascension was somewhat less,
extending from 15\fh0--17\fh0, and 17\fh3--20\fh0.

Except for \S\ref{sec:counterparts}, all results reported in this Letter
are based on the MSAM1-92 observations.


\section{Data Analysis}

The initial data reduction used in this analysis has been reported 
in Papers I and II; only the salient features are reviewed 
here.  Note that this Letter is based on the reanalyzed version of the 
MSAM1-92 data including the full instrument covariance matrix 
(\cite{inman96}).
Two nearly statistically independent demodulations are formed from the data.  
The single-difference demodulation subtracts the right beam from the 
left beam, and ignores the central beam data; the double-difference 
demodulation subtracts the average of the left$+$right beams from the 
central beam.  
The single-difference demodulation has an antisymmetric two-lobe
response pattern, while the double-difference demodulation has a 
symmetric three-lobe pattern.
The data are binned by position and angular orientation of the 
demodulated antenna pattern on the sky.  
To remove instrumental drift, a slowly varying function of time is 
fit simultaneously with the bins.
The bin size is 0\fdg12 in position, and 10\arcdeg\ in angular orientation.  

The dominant sources of brightness contrast expected in our bands are
anisotropy in the CMBR, and thermal emission from warm ($\sim 20$K) 
interstellar dust (\cite{weiss80}, \cite{bennett92}).  
Before searching for compact features, we use one of our spectral degrees
of freedom to make a simple correction for the presence of dust 
emission.  Unlike our previous analyses for CMBR anisotropy, here we do not 
impose a full spectral model on the data.
Rather, after a minimal correction for dust contamination, we model the 
spatial structure of several bright, compact features, and study the spectral 
nature of those features.

\subsection{Foreground Dust Subtraction}

We estimate and subtract the contribution of foreground interstellar
dust emission using the 22.6~\icm\ band, assuming a $\nu^{1.5}$ 
emissivity Rayleigh-Jeans spectrum (the noise estimates in the remaining 
three bands are adjusted to reflect this subtraction).
In making this correction, we assume that the 22.6~\icm\ band is 
an uncontaminated monitor of dust.  
Varying the exponent of the power law emissivity of the dust by $\pm 0.2$
has a negligible effect on the final results, compared with the statistical 
uncertainties.  The dust correction is most important for the 16.5~\icm\
band, and has little effect at the lower frequencies.

\subsection{Search for Unresolved Features}
\label{sec:search}

After removing the estimated dust contamination, 
we search the remaining 3 frequency bands for unresolved features using 
an $F$--statistic to measure significance.  
Each band is searched independently, and then the source lists are compared.  
Since the telescope is scanned at roughly constant declination, 
we parameterize the location of the features
as a function only of right ascension, and constrain the central declination 
of the feature to follow a slow linear function of right ascension.  

\begin{figure}[tbhp] 
\plotone{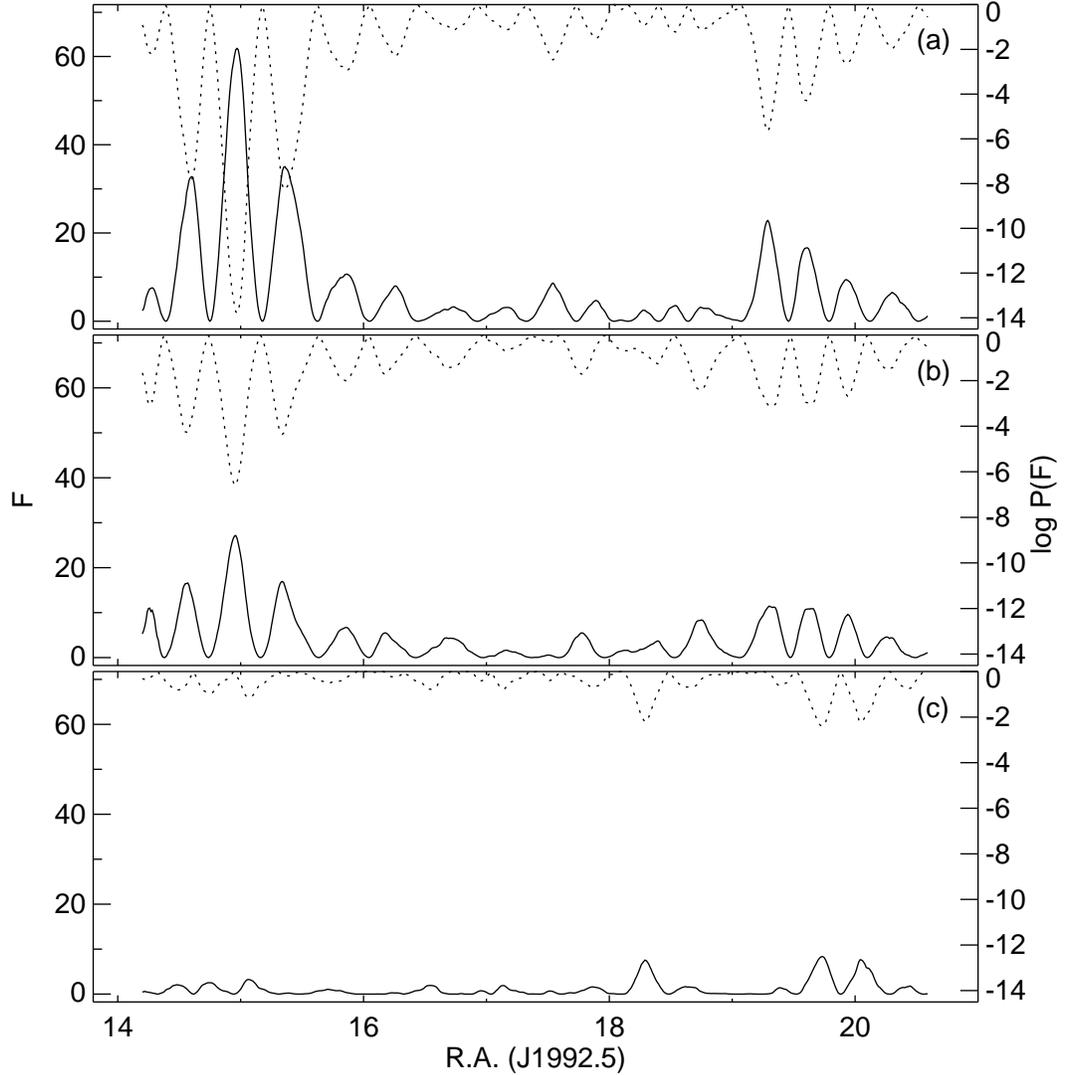} \\[3ex]
\caption[ ]
  {$F$ Statistic (solid curve) and 
   cumulative probability $P(F)$ (dotted curve); 
  (a) 5.7~\icm\ band, (b) 9.0~\icm\ band, (c) 16.5~\icm\ band.  Low values
   of $P(F)$ correspond to significant detections.  Note that the multi-lobe
   antenna patterns cause individual sources to alias into several locations
   (see text).
  \label{fig:significance}
}
\end{figure}

We begin by calculating the raw $\chi^2$ of the data in each band,
\[ \chi^2_0 = \sum_{ij} (\rawi-O_i) \wgt (\rawj-O_j), \]
where $\rawi$ are the binned data with associated noise covariance $\covar$,
and the sum is over both demodulations as well as bins.  
$O_i$ is one of four arbitrary offset parameters
corresponding to the two demodulations and two portions of the flight,
appropriate to datum $i$.
Then, as a function of right ascension $\alpha$, 
we find the flux density $S_\nu(\alpha)$ that minimizes
\[
\chi^2_S(\alpha) = \sum_{ij}\left[ \rawi - O_i - S_\nu(\alpha) \beami \right]
                             \wgt
                            \left[ \rawj - O_j - S_\nu(\alpha) \beamj \right],
\]
where $\beami$ is the value of the beam map (single or double difference, as
appropriate) centered on observation point $i$, evaluated at right 
ascension $\alpha$.
The significance of the fitted flux density is determined 
by forming the statistic,
$ F = ( \chi^2_0 - \chi^2_S ) / ( \chi^2_S/N), $
where $N$ is the number of degrees of freedom remaining in the data after 
fitting out the source.  
This statistic should be distributed according 
to an $F$-distribution with 1 and $N$ degrees of freedom (\cite{martin71}),  
which allows us
to calculate the cumulative probability $P(F)$ of drawing a value greater 
than or equal to $F$ if the null hypothesis---that the actual value 
$S_\nu(\alpha)=0$---were true. 
Values of $F \gg 1$ are highly unlikely under the null hypothesis,
and correspond to significant detections of compact features in the data.
Figure~\ref{fig:significance} shows $F(\alpha)$ and $P(F)$ for each of the 
three frequency bands. 

Since we are plotting the cumulative probability of the null hypothesis
being true, lower values of $P(F)$ correspond to more significant detections. 
We have multiple lobe beam patterns, so a bright unresolved feature will 
alias to several values of right ascension separated by the 40\arcmin\
beam throw.  Thus, the clusters of 
peaks near 15\hour\ and 19\fh3 do not correspond to multiple features, but 
rather the aliasing of a single feature into the side beams.

We build a list of candidate unresolved features for each band by iteratively
searching for and subtracting the most significant feature,  
defined here as the maximum value of $F(\alpha)$.
After each feature is subtracted, $S_\nu(\alpha)$, and $F(\alpha)$ 
are re-evaluated to help suppress the aliasing in right ascension.
We continue this procedure until $F_{\rm max} \lesssim 11$, corresponding to
$P(F) \ge 10^{-3}$. Three candidate features were identified this
way for the 5.7 and 9.3 \icm\ bands.
We allow each feature to be positive or negative, but the three brightest
features are all positive.
No features were found at 16.5 \icm\ which satisfy the $P(F) < 10^{-3}$ 
requirement.  Table~\ref{table:clean} lists the results of this search, 
along with the corresponding central declinations.

\begin{deluxetable}{rrrrr}
\tablewidth{0pt}
\tablecaption{Candidate Unresolved Features \label{table:clean}}
\tablehead{
  \colhead{R.A.\tablenotemark{a}} &
  \colhead{Dec\tablenotemark{a}} & \colhead{$F$} & \colhead{$P(F)$} &
  \colhead{$S_\nu$ (Jy)} 
}
\startdata
\cutinhead{--- Band 1 (5.7 \icm) ---}
14\fh97 & $+$81\fdg97 & 62 & $1.6 \times 10^{-14}$     & $+5.5 \pm 0.6$ \nl
19\fh29 & $+$81\fdg87 & 25 & $6.9 \times 10^{-7\phn}$  & $+2.5 \pm 0.5$ \nl
19\fh92 & $+$81\fdg85 & 14 & $1.6 \times 10^{-4\phn}$  & $+1.9 \pm 0.5$ \nl
\cutinhead{--- Band 2 (9.3 \icm) ---}
14\fh96 & $+$81\fdg97 & 27 & $2.5 \times 10^{-7\phn}$  & $+4.9 \pm 0.9$ \nl
19\fh30 & $+$81\fdg87 & 12 & $5.8 \times 10^{-4\phn}$  & $+2.7 \pm 0.8$ \nl
19\fh94 & $+$81\fdg85 & 12 & $4.9 \times 10^{-4\phn}$  & $+2.9 \pm 0.8$ \nl
\enddata
\tablenotetext{a}{J1992.5 coordinates.}
\end{deluxetable}

\subsection{Spatial Extent}
\label{sec:fits}

To determine the spatial extent of each feature, we fit the data to a Gaussian 
shaped surface brightness model,
$I_\nu e^{-(\vec{x}-\vec{x}_\alpha)^2/2 \sigma^2}$, with the central
brightness $I_\nu$, right ascension $\alpha$, and width 
$\sigma=\fwhm/(2\sqrt{2 \ln 2})$ as free parameters.  
The fits are performed using the Levenberg-Marquardt method (\cite{press92}).
The total flux density of the feature is related to the central
brightness by $S_\nu = 2\pi \sigma^2 I_\nu$.  

The fits near 15\hour\ indicate a clearly resolved feature in both the 
5.7 and 9.3 \icm\ bands, with $\fwhm \approx 35 \arcmin$.
The results of these fits are given in Table~\ref{table:fit1}.
There are substantial correlations between \fwhm\ and $I_\nu$, so
these are also listed.  From $I_\nu$, we can also infer an equivalent
peak $\Delta T_{\rm CMBR} = 360\pm50 \mu$K (5.7 \icm), $330\pm100 \mu$K 
(9.3 \icm), near 15\hour.
The fits near 19\fh3 and 19\fh9 do not show evidence for significant
spatial extent.  We can infer 68\% (1--$\sigma$) upper bounds 
on \fwhm\ from the estimated parameter variance of the fit; 
these give $\fwhm < 25\arcmin$ at 19\fh3 and $\fwhm < 35\arcmin$ at 19\fh9,
consistent in both bands.

Because of the close proximity of the 19\fh3 and 19\fh9 features, there is
some correlation between their measured flux densities.  To account for 
this, the fits are repeated, omitting \fwhm\ while simultaneously 
fitting for $\alpha$ and $S_\nu$ of both features.  
The results of these fits are shown in Table~\ref{table:fit2}.

\begin{deluxetable}{rlrr}
\tablecolumns{4}
\tablewidth{0pt}
\tablecaption{Fit Results at 15\hour \label{table:fit1}}
\tablehead{
  \colhead{Parameter}&\colhead{Fit Result\tablenotemark{a}}
                      &\multicolumn{2}{c}{Correlations} \\
  \cline{3-4}\\
  \colhead{} & \colhead{} & \colhead{$\alpha$} & \colhead{\fwhm}
          }
\startdata
\sidehead{5.7\icm \quad ($\chi^2=720/627$)}
  $\alpha:~$ &   14\fh96$\pm$0\fh02    & \nodata & \nodata \nl
     \fwhm:~ & 36\arcmin$\pm$8\arcmin  & $-0.35$ & \nodata \nl
 $I_{5.7}$:~ &       158$\pm$24 kJy/sr & $+0.21$ & $-0.68$ \nl
\tableline
\sidehead{9.3\icm \quad ($\chi^2 = 646 / 627$)}
  $\alpha$:~ &   14\fh95$\pm$0\fh03    & \nodata & \nodata \nl
  \fwhm:~    & 33\arcmin$\pm$13\arcmin & $-0.32$ & \nodata \nl
 $I_{9.3}$:~ &       137$\pm$40 kJy/sr & $+0.22$ & $-0.79$ \nl
\tableline
\sidehead{16.5\icm \quad ($\chi^2 = 542 / 629$)}
 $I_{16.5}$:~ & $-27\pm$77 kJy/sr & \nl
\enddata
\tablenotetext{a}{Errors are statistical.}
\end{deluxetable}

\begin{deluxetable}{rlrrr}
\tablecolumns{5}
\tablewidth{0pt}
\tablecaption{Fit Results at 19\fh3 and 19\fh9 \label{table:fit2}}
\tablehead{
  \colhead{Parameter}&\colhead{Fit Result\tablenotemark{a}}
                        &\multicolumn{3}{c}{Correlations} \\
  \cline{3-5}\\
  \colhead{} & \colhead{} & \colhead{$\alpha^{\rm 19\fh3}$} & 
               \colhead{$\alpha^{\rm 19\fh9}$} & \colhead{$S_\nu^{\rm 19\fh3}$}
          }
\startdata
\sidehead{5.7\icm \quad ($\chi^2=760/626$)}
  $\alpha^{\rm 19\fh3}$:~ &   19\fh29$\pm$0\fh02    
                                       & \nodata & \nodata & \nodata \nl
  $\alpha^{\rm 19\fh9}$:~ &   19\fh92$\pm$0\fh03    
                                       & $-0.04$ & \nodata & \nodata \nl
 $S_{5.7}^{\rm 19\fh3}$:~ &   2.7$\pm$0.5 Jy
                                       & $+0.06$ & $-0.06$ & \nodata \nl
 $S_{5.7}^{\rm 19\fh9}$:~ &   1.9$\pm$0.5 Jy
                                       & $+0.09$ & $+0.02$ & $+0.10$ \nl
\tableline
\sidehead{9.3\icm \quad ($\chi^2=652/626$)}
  $\alpha^{\rm 19\fh3}$:~ &   19\fh32$\pm$0\fh03    
                                       & \nodata & \nodata & \nodata \nl
  $\alpha^{\rm 19\fh9}$:~ &   19\fh93$\pm$0\fh03    
                                       & $-0.12$ & \nodata & \nodata \nl
 $S_{9.3}^{\rm 19\fh3}$:~ &   3.1$\pm$0.8 Jy
                                       & $+0.07$ & $-0.10$ & \nodata \nl
 $S_{9.3}^{\rm 19\fh9}$:~ &   3.0$\pm$0.8 Jy
                                       & $+0.08$ & $+0.06$ & $+0.11$ \nl
\tableline
\sidehead{16.5\icm \quad ($\chi^2 = 541 / 628$)}
 $S_{16.5}^{\rm 19\fh3}$:~ &   1.0$\pm$2.5 Jy
                                       & \nodata & \nodata & \nodata \nl
 $S_{16.5}^{\rm 19\fh9}$:~ &   2.9$\pm$2.6 Jy
                                       & \nodata & \nodata & $+0.07$ \nl
\enddata
\tablenotetext{a}{Errors are statistical.}
\end{deluxetable}

\subsection{Limits at 16.5 \icm}

While there are no bright features detected at 16.5 \icm, we can find
limits for the corresponding features found at the lower frequencies.  For the
first feature, we fit using our extended model, fixing $\alpha=14\fh95$ and 
$\fwhm=35\arcmin$, while for the other features, we fit assuming 
completely unresolved sources with $\alpha=19\fh31$ and 
$\alpha=19\fh92$ fixed.  
In all cases, the results are consistent with zero flux, and upper 
bounds can be inferred from the variance.  
These are also listed in Tables \ref{table:fit1} and \ref{table:fit2}

\section{Spectral Models for Features}

\begin{figure} 
\plotone{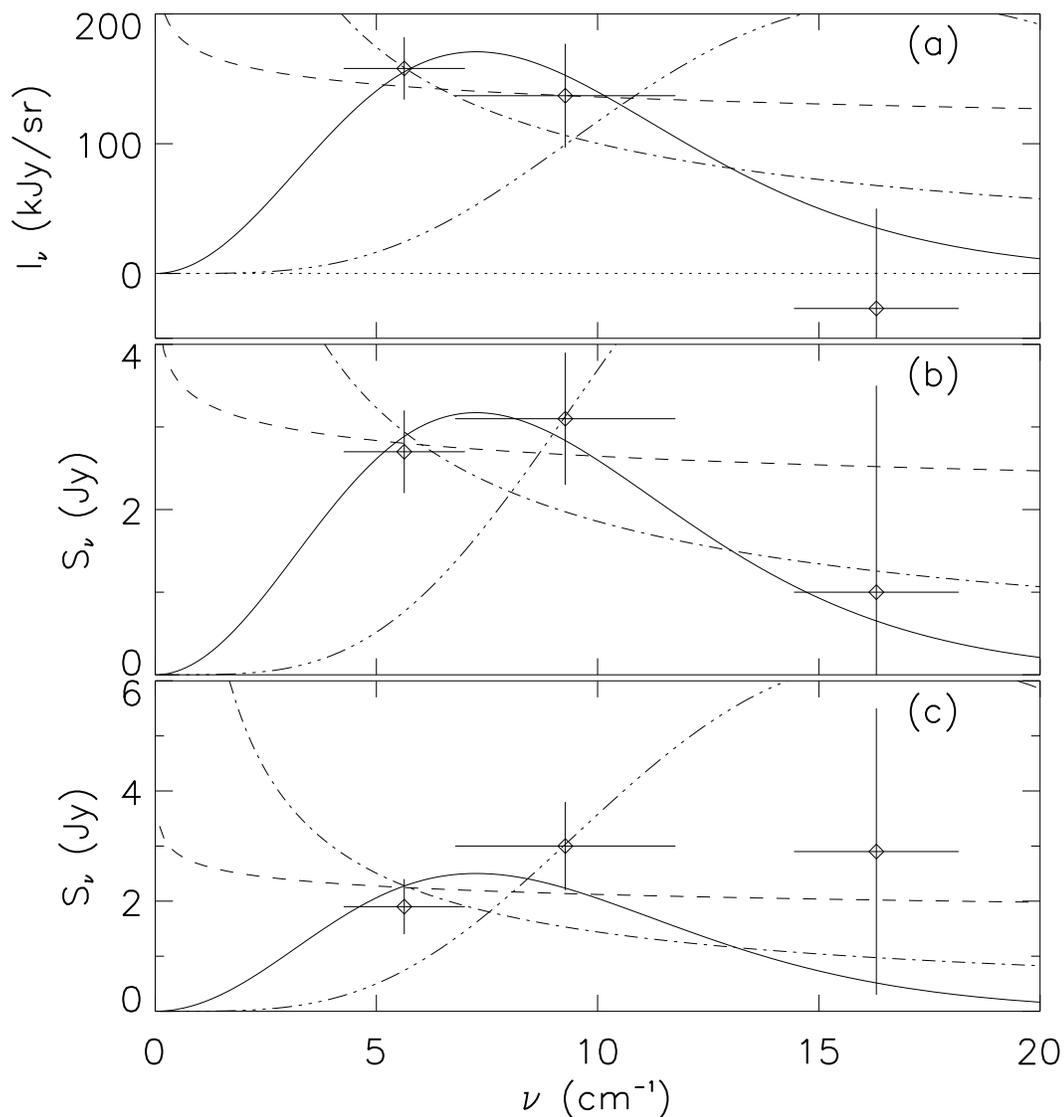} \\[5ex]
\caption[ ]
  {Measured spectra for the three features; 
   (a) 14\fh95, (b) 19\fh3, and (c) 19\fh9.  Horizontal bars show the effective
   width of each band.
   Superimposed are the best fit curves for a CMBR anisotropy spectrum 
   (solid), free-free  (dashed), synchrotron emission (dot-dashed), and
   cold dust (triple-dot-dashed). 
   \label{fig:spectra}
}
\end{figure}

We now compare the measured spectra of the features with four
simple models for astrophysical sources:
CMBR anisotropy, free-free emission with $I_\nu \propto \nu^{-0.1}$,
synchrotron emission with $I_\nu \propto \nu^{-0.8}$, and cold dust
emission with 
$I_\nu \propto \nu^2 \times (B_\nu(4.8{\rm K}) - B_\nu(T_{\rm CMBR}))$.
The last of these is motivated by the possible detection of a cold dust
contribution to the Galactic spectrum observed by the \COBE/FIRAS experiment
(\cite{wright91}, \cite{reach95}).
Each of the models is fit to the values listed in Tables \ref{table:fit1}
and \ref{table:fit2}, the results of which are displayed in 
Figure~\ref{fig:spectra}.  
The $\chi^2$ for the spectral fits are in Table~\ref{table:spectra}.
The relative calibration uncertainty is small compared to the statistical
uncertainties, and is not included in the $\chi^2$ values.

\begin{deluxetable}{ccccc}
\tablecolumns{5}
\tablewidth{0pt}
\tablecaption{Spectral Fit $\chi^2$ Values\label{table:spectra}}
\tablehead{
  \colhead{}     & \colhead{}     & \colhead{Free-} & \colhead{Sync} &
                   \colhead{Cold} \\
  \colhead{R.A.} & \colhead{CMBR} & \colhead{Free} & \colhead{} &
                   \colhead{Dust} 
          }
\startdata
15\fh0 & 0.8 & 4.5 & 2.0 & 40. \nl
19\fh3 & 0.3 & 0.7 & 2.0 & 20. \nl
19\fh9 & 2.4 & 1.7 & 4.3 & 6.9 \nl
\tablecomments{All $\chi^2$ are for 2 degrees of freedom. \\
\vspace{1in} }
\enddata
\end{deluxetable}

All three features are in agreement with the CMBR model.
The first two are significantly discrepant with a cold dust spectrum,
while the feature near 19\fh9 is marginally consistent with that spectrum.  
Furthermore, this location corresponds to the brightest warm dust emission
in our field.  Based on this, we cannot rule out the possibility that 
the 19\fh9 feature is be due to galactic dust emission.

From these data alone, either free-free or synchrotron emission 
gives an acceptable fit, and could explain the detections. 
This would, however, require implausibly large amplitudes for emission
at lower frequencies.  
For example, the best fit free-free spectra, extrapolated to 40~GHz
(Q-band), imply antenna temperatures of 3.4, 1.1, and 0.9~mK for the
three features; measurements made at Saskatoon of the same field 
strongly rule out such bright signals (\cite{netterfield96}).
The limits on synchrotron emission are even stronger, since the extrapolated
temperatures are roughly 3 times greater.
Recent measurements of H$\alpha$ emission in this part of the sky imply
the free-free contribution to the MSAM1 signal at 5.7\icm\ is 
$<$500~Jy/sr rms (\cite{simonetti96b}), several orders of magnitude smaller
than the signal we report.

\section{Searches for Counterparts}
\label{sec:counterparts}

\subsection{Catalog Searches}
We consider the possibility that these detections correspond to 
compact radio (or infrared) point sources.  Note that
the power law models corresponding to free-free and synchrotron also can
be interpreted as flat-spectrum or steep-spectrum radio spectra,
respectively, typical of extragalactic radio sources.
No counterparts for the features were found within one beamwidth of the
feature locations in the 5~GHz S5 Polar Cap Survey (\cite{kuehr81}).
For all three features, the best-fit flat
spectra predict 5~GHz fluxes greater than 3~Jy, while the best-fit
steep spectra predict fluxes greater than 40 Jy. The S5 survey is
estimated to be complete at 5~GHz down to 250 mJy, providing a strong
limit on any flat- or steep-spectrum radio counterpart to the detections.

A similar search was performed with the \IRAS\ 100 \micron\ Faint Source
Catalog (\cite{moshir92}).  
A total of 5 objects ranging in flux from 0.9--2.6~Jy were 
found within one beamwidth of any one of the three features, 
consistent with the average \IRAS\ FSC source density of $\sim 4/$sq. degree.  
The contribution of these sources to the MSAM signal is shown to be
negligible using the 1.25~mm to 100~\micron\ flux ratio, 
$(3.1 \pm 0.5)\times 10^{-3}$, determined from a complete sample of \IRAS\ 
galaxies, together with a relatively flat $\nu^{3.2}$ spectrum 
locally around 1.25~mm (\cite{franceschini95}).  
The greatest contribution is at 16.5~\icm, where it is estimated to be
$\sim 0.1$~Jy for each of the latter two features, and $\sim 50$~mJy for
the 15\hour\ feature.  The contributions at 9.3~\icm\ are a factor of 5
smaller than these, while at 5.7~\icm\ they are a factor of 30 smaller.
These estimates are reinforced with direct observations at 90~GHz of known IRAS
sources in the vicinity of the 15\hour\ feature, none of which show 
significant flux density above the rms noise of 25--50~mJy (\cite{chernin94}).

\subsection{MSAM1-94}
In 1994, we flew the MSAM instrument a second time to observe the same
field as the 1992 flight.  While the sky overlap between MSAM1-94 and MSAM1-92
was not perfect, it was sufficient to permit an overall
comparison of the measurements (\cite{inman96}).  
We have repeated the fits of \S\ref{sec:fits} on the MSAM1-94 data to see if
the bright features are present.  Near 15\hour, there is a corresponding
hot spot at $\alpha = 14\fh85 \pm 0\fh03$ with 
$\fwhm = 38\arcmin \pm 12 \arcmin$, $I_{5.7} = 106 \pm 32$~kJy/sr, and
$I_{9.3} = 71 \pm 62$~kJy/sr (statistical errors only).
Near 19\fh3 and 19\fh9, there are corresponding features that are 
fit simultaneously as before.  At 5.7~\icm, the resulting parameters are 
$\alpha=19\fh31 \pm 0\fh02$, $S_{5.7}^{19\fh3} = 1.3 \pm 0.3$~Jy, and
$\alpha=19\fh93 \pm 0\fh02$, $S_{5.7}^{19\fh9} = 1.4 \pm 0.4$~Jy.  
At 9.3~\icm, the fit gives
$\alpha=19\fh26 \pm 0\fh03$, $S_{9.3}^{19\fh3} = 1.8 \pm 0.5$~Jy, and
$\alpha=19\fh93 \pm 0\fh02$, $S_{9.3}^{19\fh9} = 3.2 \pm 0.7$~Jy. 
The correlations in the MSAM1-94 fits are larger than the MSAM1-92 fits;
in particular, the 19\fh3 and 19\fh9 flux densities have variance correlations 
of $+0.49$ at 5.7~\icm, and $+0.42$ at 9.3~\icm.
The somewhat smaller flux densities in MSAM1-94 are consistent with the 
pointing differences between the two flights.

\subsection{Other Observations}
Since the initial report of these detections in Paper~I, a number
of groups have searched one or more of these regions for corresponding 
sources at different wavelengths and/or resolutions.  Of particular note
is the search reported by \cite{church95} using the Caltech
Submillimeter Observatory on Mauna Kea.  This  search, which made a deep
map of the sky around the 15\hour\ feature at 4.7\icm\ using a 1\farcm7 beam,
places an upper limit on any point source  in that field of $S_{4.7}<1$~Jy.

Another observation of this field has recently been reported
by \cite{netterfield96}.  These observations were made from the
ground at Saskatoon, Saskatchewan, Canada, in Q-band 
(36--46 GHz, 1.2--1.5 \icm) at 0\fdg5 resolution, a portion of which
were designed to reproduce, as closely as possible, the MSAM observations.  
A three-lobe beammap, similar to the MSAM double-difference demodulation, 
was synthesized (the single-difference demodulation is not usable from the 
ground due to atmospheric noise).  
Although having somewhat less statistical weight than the MSAM observations, 
the Saskatoon results correlate well with MSAM, including the 
bright feature near 15\hour\ right ascension.  

\section{Simulations}
\label{sec:simulations}

The presence of bright, compact features in the MSAM1-92 data was unexpected.
In Paper~I, we questioned whether CMBR fluctuations obeying Gaussian
statistics could produce such features.  To address this issue, we have 
performed a detailed Monte Carlo simulation of the MSAM observations.

The approach used is similar to that outlined in \S4.3 of 
Paper~I\footnote{Eq (2) in Paper~I contains a typographical error, and should 
read $\left< s_k s_l \right> = 
\int d\bx_1 d\bx_2 H_k(\bx_1)H_l(\bx_2)C(|\bx_1-\bx_2|)$.}.
We generate numerical realizations of the observations, including 
contributions from instrument noise and CMBR temperature anisotropies, 
assuming the anisotropy is a Gaussian random field described by a two-point 
correlation function, $C(\theta)$.  In Paper~I, we set limits on the overall
amplitude of $\Delta T/T = \sqrt{C(0)}$ by using the likelihood ratio 
statistic $\lambda$; here, we use those upper and lower bounds to 
generate realizations which are then searched according to the procedure 
of \S\ref{sec:search} for unresolved features.
Unlike Paper~I, here we simulate both demodulations simultaneously, 
using $\theta_c=0\fdg4$.
We thus determine the distribution of $n$, the number of unresolved 
features with $P(F) \le 10^{-3}$, corresponding to the range of $\Delta T/T$
fluctuations found in the data.  
Our measured number of sources is $n = 2$ or 3,
depending on the assignment of the 19\fh9 feature to CMBR or galactic 
dust.  For $\Delta T/T=1\times10^{-5}$, we find the cumulative probability 
$P(n \ge 2) = 0.20$, $P(n \ge 3) = 0.04$, while for 
$\Delta T/T=3\times10^{-5}$, $P(n \ge 2) = 0.90$, $P(n \ge 3) = 0.69$.
From this, we conclude that bright, compact features such as those 
observed are in fact consistent with our assumption of Gaussian statistics.

We have also studied the distribution of $n$ using a more physically
motivated correlation function, calculated for ``standard'' flat cold dark
matter (CDM) with $\Omega=1, \Omega_B=0.05, \Lambda=0, 
H_0=50 {\rm km}/{\rm s}/{\rm Mpc}$, and no reionization (\cite{sugiyama95}).  
Using this model, normalized to the \COBE/DMR detection 
$Q_{\rm rms-PS}=20 \mu$K (\cite{gorski94}),  we find reasonable probabilities
for the observed $n$: $P(n \ge 2) = 0.43$, $P(n \ge 3) = 0.15$.  
Thus, the detections are consistent with adiabatic fluctuation models such as
CDM.  Similar results have also been reported for a more generic study of
0\fdg5 resolution CMBR maps (\cite{kogut94}).

\section{Conclusions} 
All three features are best described 
as temperature anisotropies in the CMBR, although the possibility remains
that one of these is due to galactic foreground dust emission.  While we were
initially surprised to find such bright, compact features in the data,
we now understand that these are consistent with our assumption
of Gaussian fluctuations.  Although our catalog searches were all null, 
we find the most compelling argument that the features are not true
point-sources comes from the direct search by \cite{church95}.
Finally, the morphological agreement reported by the Saskatoon experiment, 
consistent with a CMBR anisotropy spectrum spanning three 
octaves in frequency from Q-band (1.2 \icm) to the MSAM 9.3 \icm\ band, 
leaves little room for alternate explanations of the detection.
Given the agreement with a CMBR temperature anisotropy spectrum, 
together with the implausability of other explanations, we 
conclude that at least the first two, and possibly all three of these 
detections, are cosmological in origin. 

\acknowledgments
We would like to thank the staff of the National Scientific Balloon Facility
(NSBF) for their outstanding support during the two flight campaigns.
We would also like to thank E. Magnier, R. Rutledge, C. Bennett, 
R. Bond, J. Silk, and C. Lawrence for many helpful discussions of these
features over the years.
This research has made use of NASA's Astrophysics Data System abstract and 
article services and IPAC's XCATSCAN catalog scanning system, as well as 
software tools from the Free Software Foundation.
The National Aeronautics and Space Administration
supports this research through grants  NAGW 1841, RTOP 188-44, NGT 50908, 
and NGT 50720.


\begin{thebibliography}{}

\bibitem[Bennett {\em et~al.} 1992]{bennett92}
Bennett, C.~L. {\em et~al.} 1992, \apjl, {\bf 396}, L7.

\bibitem[Cheng {\em et~al.} 1994]{cheng94}
Cheng, E.~S. {\em et~al.} 1994, \apjl, {\bf 422}, L37.

\bibitem[Cheng {\em et~al.} 1996]{cheng95}
Cheng, E.~S. {\em et~al.} 1996, \apjl, {\bf 456}, L71.

\bibitem[Chernin and Scott 1995]{chernin94}
Chernin, L.~M. and Scott, D. 1995, Astronomy and Astrophysics, {\bf 296}, 609.

\bibitem[Church {\em et~al.} 1995]{church95}
Church, S.~E., Mauskopf, P.~D., Ade, P.~A.~R., Devlin, M.~J., Holzappel, W.~L.,
  Wilbanks, T.~M., and Lange, A.~E. 1995, \apjl, {\bf 440}, L33.

\bibitem[{Donat III} and Boksenberg 1993]{almanac94}
{Donat III}, W. and Boksenberg, A., editors 1993.
\newblock {\em The Astronomical Almanac for the year 1994}.
\newblock {U.S. Government Printing Office} and {HMSO}, {Washington, D.C.} and
  {London}.

\bibitem[Fixsen {\em et~al.} 1996]{fixsen96a}
Fixsen, D.~J. {\em et~al.} 1996, \apj.
\newblock submitted, preprint {\tt astro-ph/9512006}.

\bibitem[Franceschini and Andreani 1995]{franceschini95}
Franceschini, A. and Andreani, P. 1995, \apjl, {\bf 440}, L5.

\bibitem[G\'orski {\em et~al.} 1994]{gorski94}
G\'orski, K.~M. {\em et~al.} 1994, \apj, {\bf 430}, L89.

\bibitem[Griffin {\em et~al.} 1986]{griffin86}
Griffin, M.~J., Ade, P.~A.~R., Orton, G.~S., Robson, E.~I., Gear, W.~K., Nolt,
  I.~G., and Radostitz, J.~V. 1986, Icarus, {\bf 65}, 244.

\bibitem[Hagen and Boksenberg 1991]{almanac92}
Hagen, J.~B. and Boksenberg, A., editors 1991.
\newblock {\em The Astronomical Almanac for the year 1992}.
\newblock {U.S. Government Printing Office} and {HMSO}, {Washington, D.C.} and
  {London}.

\bibitem[Inman {\em et~al.} 1996]{inman96}
Inman, C.~A. {\em et~al.} 1996, \apjl.
\newblock submitted, preprint {\tt astro-ph/9603017}.

\bibitem[Kogut, Hinshaw, and Bennett 1995]{kogut94}
Kogut, A., Hinshaw, G., and Bennett, C.~L. 1995, \apjl, {\bf 441}, L5.

\bibitem[K\"uhr {\em et~al.} 1981]{kuehr81}
K\"uhr {\em et~al.} 1981, \aj, {\bf 86}, 854.

\bibitem[Martin 1971]{martin71}
Martin, B.~R. 1971.
\newblock {\em Statistics for Physicists}.
\newblock Academic, London.

\bibitem[Moshir {\em et~al.} 1992]{moshir92}
Moshir, M. {\em et~al.} 1992.
\newblock {\em IRAS Faint Source Survey: Explanatory Supplement}.
\newblock JPL D-10015 8/92. JPL, Pasadena, CA., second edition.

\bibitem[Netterfield {\em et~al.} 1996]{netterfield96}
Netterfield, C.~B., Devlin, M.~J., Jarosik, N., Page, L., and Wollack, E.~J.
  1996, \apj.
\newblock submitted, preprint {\tt astro-ph/9601197}.

\bibitem[Press {\em et~al.} 1992]{press92}
Press, W.~H., Teukolsky, S.~A., Vetterling, W.~T., and Flannery, B.~P. 1992.
\newblock {\em Numerical Recipes in {FORTRAN}: The Art of Scientific
  Computing}.
\newblock Cambridge University Press, Cambridge, 2nd edition.

\bibitem[Reach {\em et~al.} 1995]{reach95}
Reach, W. {\em et~al.} 1995, \apj, {\bf 451}, 188.

\bibitem[Simonetti, Topasna, and Dennison 1996]{simonetti96b}
Simonetti, J.~H., Topasna, G.~A., and Dennison, B. 1996, \baas.
\newblock submitted.

\bibitem[Sugiyama 1995]{sugiyama95}
Sugiyama, N. 1995, \apjs, {\bf 100}, 281.

\bibitem[Weiss 1980]{weiss80}
Weiss, R. 1980, \araa, {\bf 18}, 489.

\bibitem[White, Scott, and Silk 1994]{white94}
White, M., Scott, D., and Silk, J. 1994, \araa, {\bf 32}, 319.

\bibitem[Wright {\em et~al.} 1991]{wright91}
Wright, E.~L. {\em et~al.} 1991, \apj, {\bf 381}, 200.

\end{thebibliography}

\end{document}